\documentclass[journal]{IEEEtran}
\usepackage{amsrefs}
\newenvironment{rezabib}
  {\bibdiv\biblist\setupbib}
  {\endbiblist\endbibdiv}

\def\setupbib{\catcode`@=\active}
\begingroup\lccode`~=`@
  \lowercase{\endgroup\def~}#1#{\gatherkey{#1}}
\def\gatherkey#1#2{\gatherkeyaux{#1}#2\gatherkeyaux}
\def\gatherkeyaux#1#2,#3\gatherkeyaux{\bib{#2}{#1}{#3}}
\usepackage{blindtext}
\usepackage{graphicx}
\usepackage{filecontents}
\usepackage[noadjust]{cite}
\usepackage{amsmath}
\usepackage[mathcal]{eucal}
\usepackage{amssymb} 
\usepackage[hyperindex]{hyperref}
\DeclareMathOperator*{\argmin}{argmin}

\begin{document}
%
\title{Non-Rigid Point Set Registration Networks}

\author{Lingjing~Wang,~Jianchun~Chen,~Xiang~Li
        and~Yi~Fang
\thanks{L.Wang is with MMVC Lab, the Department
of Mathematics, New York University, New York,
NY, 30332 USA e-mail: lingjing.wang@courant.nyu.edu. J.Chen is with the MMVC Lab, New York University, New York,
NY, 30332 USA e-mail: jc7009@nyu.edu. X.Li is with the MMVC Lab, New York University, New York, NY, 30332 USA e-mail: lixiang709709@gmail.com. Y.Fang is with MMVC Lab, Dept. of ECE, NYU Abu Dhabi, UAE. Dept. of ECE, NYU Tandon School of Engineering, USA. USA e-mail: yfang@nyu.edu.}
\thanks{Corresponding author. Email: yfang@nyu.edu}
}

\maketitle
\begin{abstract}
Point set registration is defined as a process to determine the spatial transformation from the source point set to the target one. Existing methods often iteratively search for the optimal geometric transformation to register a given pair of point sets, driven by minimizing a predefined alignment loss function. In contrast, the proposed point registration neural network (PR-Net) actively learns the registration pattern as a parametric function from a training dataset, consequently predict the desired geometric transformation to align a pair of point sets. PR-Net can transfer the learned knowledge (i.e. registration pattern) from registering training pairs to testing ones without additional iterative optimization. Specifically, in this paper, we develop novel techniques to learn shape descriptors from point sets that help formulate a clear correlation between source and target point sets. With the defined correlation, PR-Net tends to predict the transformation so that the source and target point sets can be statistically aligned, which in turn leads to an optimal spatial geometric registration. PR-Net achieves robust and superior performance for non-rigid registration of point sets, even in presence of Gaussian noise, outliers, and missing points, but requires much less time for registering large number of pairs. More importantly, for a new pair of point sets, PR-Net is able to directly predict the desired transformation using the learned model without repetitive iterative optimization routine. Our code is available at https://github.com/Lingjing324/PR-Net.
\end{abstract}

\section{Introduction}
\subsection{Background}
Over past decades, point set matching and registration is one of the most important computer vision tasks \cite{myronenko2007non, jian2011robust,bai2007skeleton,bai2008path,myronenko2009image,ma2016non,ma2014robust,wu2012online,ling2005deformation},
serving a widespread applications such as stereo matching, medical image registration, large-scale 3D reconstruction, 3D point cloud matching, semantic segmentation and so on \cite{klaus2006segment,maintz1998survey,besl1992method,raguram2008comparative,yuille1988computational,sonka2014image}. The point set registration is mathematically defined as a process to determine the spatial geometric transformations (i.e. rigid and non-rigid transformation) that can optimally register the source point set to the target one. The desired registration algorithm can find both rigid (i.e. rotation, reflection, and shifting) and non-rigid (i.e. dilation and stretching) transformations, as well as being robust to outliers, Gaussian point drift, data incompleteness and so on. 

To formulate the problem of point set registration, existing methods \cite{ma2014robust,myronenko2009image} often iteratively search the optimal geometric transformation to register two sets of points, driven by minimizing a predefined alignment loss function. The alignment loss is usually pre-defined as a certain type of distance metric (e.g. Euclidean distance loss) between the transformed source point set and the target one. Previous efforts \cite{ma2014robust,myronenko2009image,tam2013registration} have achieved great success in point set registration through the development of a variety of optimization algorithms and distance metrics as summarized in \cite{tam2013registration}. However these methods are often not designed to handle the real-time point set registration or to deal with a large volume dataset. This limitation is mainly contributed by the fact that, for each given pair of point sets, the iterative method needs to start over a new iterative optimization process even for the trivial similar cases. This observation suggests that the existing efforts are mainly concentrated on the stand-alone development of the optimization strategies rather than the techniques to smartly transferring the registration pattern acquired from aligning one pair to another. This triggers the motivation to develop our proposed PR-Net with the hope to actively learn the registration pattern from a set of training data, consequently, to adaptively utilize that knowledge to directly predict the geometric transformation for a new pair of unseen point sets. As a result, PR-Net is capable of handling the real-time point set registration or a large volume datasets with a similar pattern. To better understand the point set registrations, we briefly review related works as follows.
\subsection{Related Works}
\noindent \textbf{Iterative registration methods.} Current mainstream point set registration methods focus on the development of optimization algorithms to estimate the rigid or non-rigid geometric transformations in an iterative routine. With the assumption that a pair of point sets are related by a rigid transformation, a registration approach is to estimate the best translation and rotation parameters in the iterative search routine aiming to minimize a distance metric between two sets of points. One of the most popular methods for rigid registration, the Iterative Closest Point (ICP) algorithm \cite{besl1992method}, was proposed to handle point set registration with least-squares estimation of transformation parameters. ICP starts with an initial estimation of rigid transformation, followed by iteratively refining the transformation by alternately choosing corresponding points from the point sets as estimate transformation parameters. The ICP algorithm is reported to be vulnerable to the selection of corresponding points for initial transformation estimation, and also incapable of dealing with non-rigid transformation.

To accommodate the deformation (e.g. morphing, articulation) between a pair of point sets, many efforts were spent in the development of algorithms to address the challenges of a non-rigid transformation. Chui and Rangarajan \cite{chui2000new} proposed a robust method to model non-rigid transformation named as thin-plate spline \cite{bookstein1989principal}. They proposed TPS-RSM algorithm with penalization on second order derivatives to optimize the parameters of the desired transformation. Ma et al. \cite{ma2015robust} introduced a $L_2E$ estimator for non-rigid registration for handling significant scale changes and rotations. In addition, Myronenko et al. \cite{myronenko2007non} proposed non-parametric coherence point drift (CPD) algorithm which leverages Gaussian mixture likelihood and penalizes derivatives of all orders of the velocity field to enforce velocity coherence so that centroids of source point set move coherently to target point set. They reported that their algorithm can be easily extended to N-dimensional space compared to TPS-RSM algorithm. Ma et al. \cite{ma2014robust} proposed a non-parametric vector field consensus algorithm to establish the robust correspondence between two sets of points. Their experimental result demonstrated that the proposed method is quite robust to outliers. In \cite{ma2016non}, the authors emphasized the importance to preserve local and global structures for non-rigid point set registration. Wang et al. \cite{wang2016path} proposed path following strategy for graph matching in order to improve the computation efficency. Zhou et al. \cite{zhou2015multi} proposed a fast alternating minimization algorithm for multi-image matching.  Existing methods have achieved great success for both rigid and non-rigid point set registration over past decades. However, they are mainly concentrated on the stand-alone development of the optimization strategies for point set registration rather than the techniques to learn the registration process as a pattern. In this paper, the deficiency of these current algorithms drives us to develop a learning-based registration paradigm that is able to actively learn the knowledge about how to register two point sets, consequently, to adaptively utilize those knowledge to directly predict the geometric transformation without the necessary to start over a new iterative search process for each similar case.

\noindent \textbf{Learning-based registration methods.} Recent great success of deep learning in various computer vision fields \cite{su2015multi,sharma2016vconv,maturana2015voxnet,qi2017pointnet,verma2018feastnet,masci2015geodesic,zeng20173dmatch} motivates researchers to start modeling the registration problem using deep neural networks \cite{rocco2017convolutional,balakrishnan2018unsupervised,zeng20173dmatch,qi2017pointnet,verma2018feastnet,masci2015geodesic}. Earlier attempt in this direction is mainly concentrated on the development of learning-based registration methods for pairwise image registration. For example, Rocco et al. \cite{rocco2017convolutional} developed a CNN architecture to predict both rigid and non-rigid transformation for 2D image matching. Balakrishnan et al. \cite{balakrishnan2018unsupervised} proposed a deep learning method to predict the non-rigid deformation field with application in deformable medical image registration. Both works share the common use of deep learning for visual feature learning from image to formulate the pairwise image correlations. The method presented in \cite{rocco2017convolutional} tends to predict the parameters of TPS-based transformation function for pairwise image registration, while the authors in \cite{balakrishnan2018unsupervised} aim to predict a smooth registration field to approximate non-rigid transformation. Though it is not a direct registration model, Zeng et al. \cite{zeng20173dmatch} proposed a volumetric 3D-CNN to learn local shape descriptor geometric patch matching. The aforementioned learning-based registration methods, despite not working on point set registration, are encouraging for us to take a further step in this paper to investigate the possibility of learning point set registration using deep neural networks. We will briefly describe our proposed PR-Net in the subsection below, and the technical details are discussed in the approach section.

\begin{figure*}
\centering
\includegraphics[height=8.8cm, width=14.3cm]{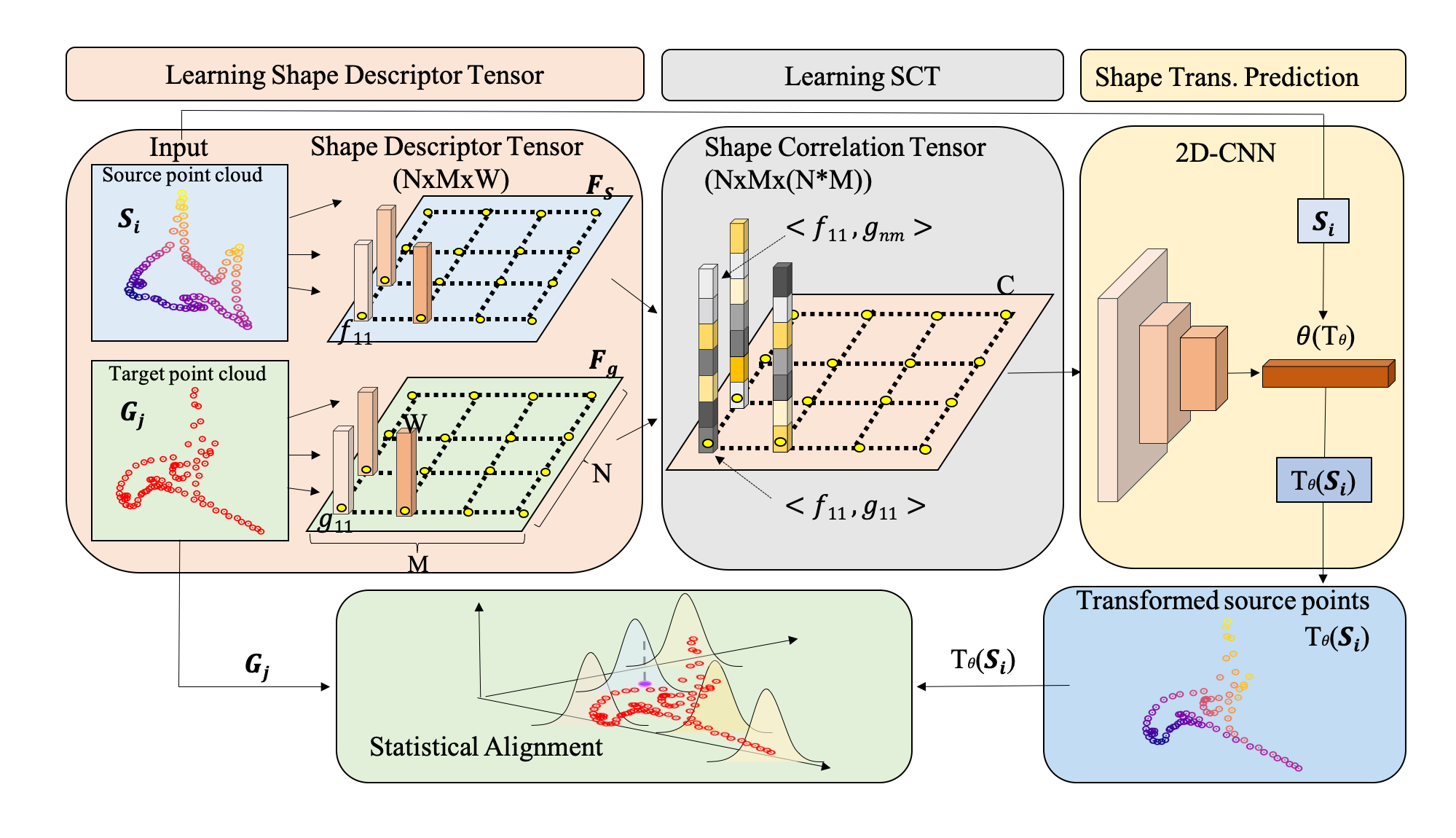}
\caption{PR-Net pipeline. The proposed PR-Net includes three parts: learning shape descriptor tensor (SCT), learning correlation tensor, and shape transformation prediction. For a pair of source point set $\mathbf{S_i}$ and target point set $\mathbf{G_j}$, we first generate two reference grids and map points of source and target point sets on them as two shape descriptor tensor $\mathbf{F_{s}}$ and $\mathbf{F_{g}}$. We define the shape correlation tensor $\mathbf{C}$ between the source and target shape descriptor tensors. By leveraging 2D-CNN, we learn the desired parameters $\theta$ of transformation $T_{\theta}$ based on the shape correlation tensor. The learned optimal model transforms source point set to be statistically aligned with the target point set.}
\label{fig222}
\end{figure*}

\subsection{Our Solution: Point Set Registration Neural Network (PR-Net)}
Different from image data with a regular grid, point cloud data is often recorded in an irregular and disordered format. Learning the point set registration requires the deep neural networks to be applicable to irregular point cloud data. In addition, unlike the image containing rich texture and color information, the point cloud is solely represented with geometric information (i.e. coordinates, curvature, normal). This suggests that a learning-based solution for point set registration needs to address two main technical challenges: 1) robust learning of both local and global geometric feature from point clouds and 2) robust learning of the transformation from  well-defined correlation measure between pairwise geometric feature sets. Therefore, the proposed PR-Net investigates two major research problems: 1) the design of the techniques for point cloud learning by introducing a novel reference operator to enable formulating the correlation measure on arbitrary-structured data, and 2) the development of learning paradigm for the geometric transformation learning from pairwise feature sets. 

Figure \ref{fig222} illustrates the pipeline of the proposed PR-Net which is composed of three main components. The first component is ``learning shape descriptor tensor''. In this component, the proposed grid-reference structure is developed to enable feature learning and formulate the correlation relationship on arbitrary-structured data. The second component is ``learning shape correlation tensor''. In this component, the shape correlation tensor is developed as a metric to further evaluate the correlation between two shape descriptor tensors of point sets to be registered. The shape correlation tensor is formulated as ``all-to-all'' point-wise computation from the pair of shape descriptor tensors evaluated in the first component. The third component is ``learning of the parameters of transformation''. In this component, we exploit the function mapping between space of the ``shape correlation tensor'' and ``the parameters of transformation'' to determine the best geometric transformation that statistically aligns the source point cloud set and the target one. In this paper, PR-Net utilizes the CNN as functional regression model to approximate the aforementioned mapping function for the parameters learning of the desired transformation. Accordingly, the main components of the pipeline indicate the main contributions of our proposed PR-Net as follows:\\

\begin{itemize}

\item We propose a novel technique to learn the global and local shape aware ``shape descriptor tensor'' directly from the point cloud with irregular and disordered format. The shape descriptor tensor is proved to be effective and efficient in extracting the geometric shape features, even for point cloud in presence of missing points, noise, and outliers.\\

\item We propose a novel shape correlation tensor to comprehensively evaluate the correlation between two point sets to be registered.\\ 

\item We propose a novel statistical alignment loss function that drives our structure to determine the optimal geometric transformation that statistically aligns the source point cloud set and the target one.\\

\item In all, we propose a novel learning-based point set registration paradigm which learns registration patterns from training data, consequently, to adaptively utilize that knowledge to directly predict the geometric transformation for aligning a new pair of point sets, without the necessity to start over a new iterative search process. \\

\end{itemize}

In conclusion, given a large number of data set for training, PR-Net demonstrates a stable generalization ability to directly predict the desired non-rigid transformation for the unseen point clouds data even in presence of a great level of noise, missing points, and outliers. 

\section{Approach}
We introduce our approach in the following sections. In section 2 A, we state our learning-based registration problem. From section 2 B to 2 D, four successive parts are illustrated to explain each module of our method in details. Section 2 B illustrates our structure for learning shape descriptor tensor for point sets. In section 2 C, we introduce shape correlation tensor based on the learned shape descriptors. The non-rigid shape transformation prediction is introduced in section 2 D. The definition of the loss function is discussed in section 2 E and the settings of the training and model configuration are explained in section 2 F. 

\subsection{Problem statement}
Prior to discussion of our approach, we first define the point set registration task. Let the training data set $\mathbf{D}=\{(\mathbf{S_i}, \mathbf{G_j}) \text{ ,where } \mathbf{S_i}, \mathbf{G_j} \subset \mathbb{R}^N \}$. We denote $\mathbf{S_i}$ source point set and $\mathbf{G_j}$ target point set. In this paper, we mainly discuss the situation when $N=2$ and $N=3$. We assume $\forall (\mathbf{S_i}, \mathbf{G_j}) \in \mathbf{D}, \exists \theta_i, T_{\theta_i}:\mathbb{R}^N \to \mathbb{R}^N, \text{ such that },T_{\theta_i} :\mathbf{x_i} \to \mathbf{x_i'} \text{ where } \mathbf{x_i} \in \mathbf{S_i}$ and $\mathbf{x_i'} \in \mathbf{G_j}$. $T_{\theta_i}$ can be rigid or non-rigid transformation with parameters $\mathbf{\theta_i}$. For previous methods, $\mathbf{\theta_i}$ is optimized in a iterative searching process to optimally align a given target and source point sets. For our method, we assume the existence of a neural network structure $g$ with a set of all its weights $\gamma$, such that  $g_{\gamma}(\mathbf{S_i},\mathbf{G_j}) = \theta_i$. Our optimization task becomes:

\begin{equation}
\begin{split}
\mathbf{\gamma^{optimal}} =\argmin_{\gamma}[\mathbb{E}_{(\mathbf{S_i},\mathbf{G_j})\sim \mathbf{D}}[\mathcal{L}(T_{g_{\gamma}(\mathbf{S_i},\mathbf{G_j})}(\mathbf{S_i}),\mathbf{G_j})]],
\end{split}
\end{equation}\\
Therefore, for a given training set $\mathbf{D}$, our task is to optimize parameters $\mathbf{\gamma}$ instead of $\mathbf{\theta}$/$T_{\theta}$. The desired $\mathbf{\theta}$/$T_{\theta}$ is our model's output. $\mathcal{L}(\cdot)$ represents a similarity measure. 

\subsection{Learning shape descriptor tensor}
The first part of our structure is learning the shape descriptor for point sets. To address the problem of irregular format of point set, we introduce two point grids $\mathbf{M_{S_i}}$ and $\mathbf{M_{G_j}}$ as reference point sets, which are overlaid on the source point set $\mathbf{S_i}$ and the target point set $\mathbf{G_j}$ respectively.

\begin{figure}[h]
\centering
\includegraphics[width=7cm,height=5cm]{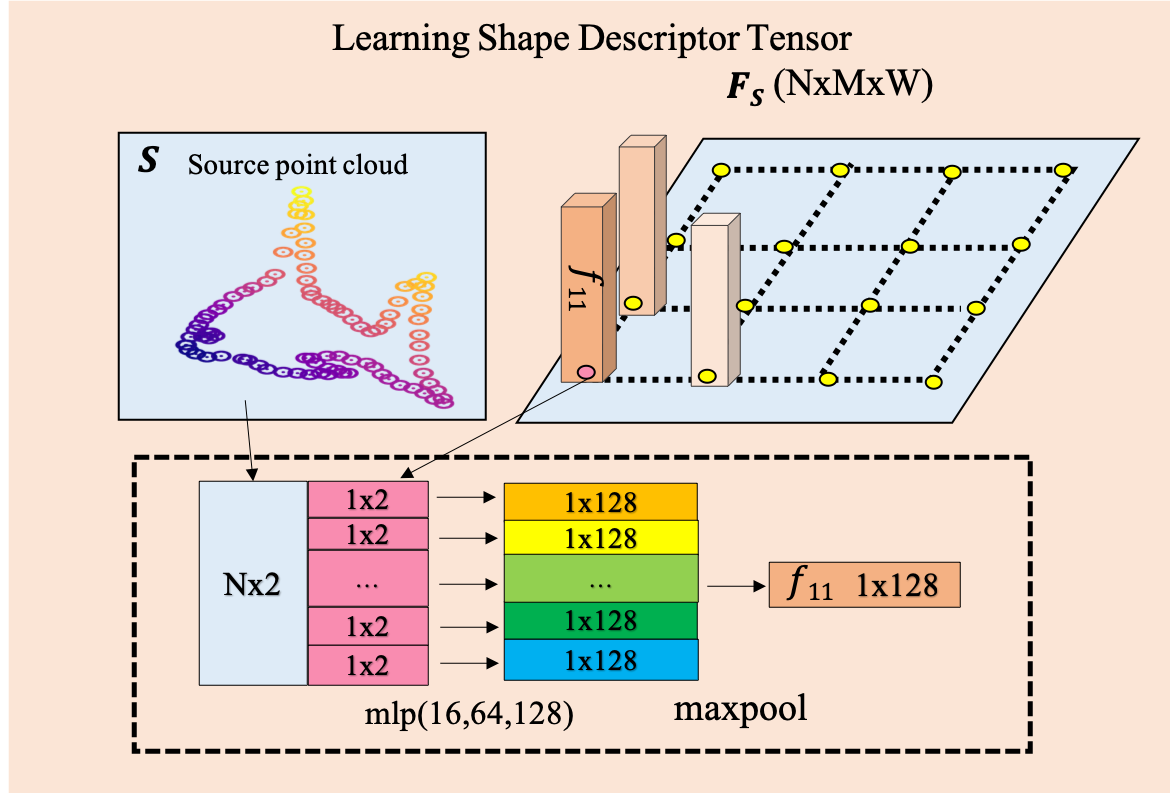}
\caption{The schema of learning shape descriptor tensor process.}
\label{fig123}
\end{figure}

For each point in the reference point sets, we learn a shape descriptor tensor $\mathbf{F_{s}^i}$ or $\mathbf{F_{g}^j}$ by mapping the local and global information of non-regular source or target point set on it. Specifically, as shown in Figure \ref{fig123}, taking $\mathbf{S_i}$ for example, $\forall \mathbf{x_i} \in \mathbf{M_{S_i}}$, $\mathbf{x_i}$ is 2D/3D geometric coordinates and we define the single layer mapping $U:(\mathbf{x_i}, \mathbf{S_i}) \to \mathbb{R}^d$ as following: 

\begin{equation}
\begin{split}
\mathcal{U}(\mathbf{x_i}, \mathbf{S_i})= \text{Maxpool} \{ \text{ReLU}(\mathbf{u_m}[\mathbf{x_i, y_i}]+\mathbf{c_m}))\}_{\mathbf{y_i}\in \mathbf{S_i}}
\end{split}
\end{equation}\\

,where parameters $\mathbf{u_m} \in \mathbb{R}^{m\times 4/6}$, $c_m \in \mathbb{R}^{m\times 1}$ and [*,*] means concatenation. For multi-layers' structure, we repeat the linear combination and Leaky-ReLU \cite{xu2015empirical} activation parts before applying the Max-pool layer. The MLP-based structure was firstly introduced in PointNet \cite{qi2017pointnet} for directly learning geometric features from point cloud. Please refer to PointNet \cite{qi2017pointnet} for more details. The single layer MLP-based function $\mathcal{U}$(*) can be regarded as a mapping to exact features from non-regular point set, which is driven by the loss function. In our case, we have three layers MLP. In this way, we transfer information of source and target point sets to two shape descriptor tensors on reference grids. We define the shape descriptor tensor $\mathbf{F_{S}^i}$ and $\mathbf{F_{G}^j}$. $\mathbf{F_{S}^i}$, $\mathbf{F_{G}^j} \in \mathbb{R}^{n\times m \times d}$ where \\

$\mathbf{F_{S}^i}$=$\begin{bmatrix}
 \mathcal{U}(\mathbf{x_{11}},\mathbf{S_i}), & \mathcal{U}(\mathbf{x_{12}}, \mathbf{S_i})& \dots & \mathcal{U}(\mathbf{x_{1n}},\mathbf{S_i}) \\
 \mathcal{U}(\mathbf{x_{21}},\mathbf{S_i}) &\mathcal{U}(\mathbf{x_{22}}, \mathbf{S_i}) & \dots &\mathcal{U}(\mathbf{x_{2n}},\mathbf{S_i}) \\
 &&\dots \\
 \mathcal{U}(\mathbf{x_{n1}},\mathbf{S_i}) &\mathcal{U}(\mathbf{x_{d2}}, \mathbf{S_i}) & \dots &\mathcal{U}(\mathbf{x_{nm}},\mathbf{S_i})
\end{bmatrix}$\\ \\

, where $\mathbf{x_{nm}} \in \mathbf{M_{S_i}}$. 
Similarly, we have the shape descriptor tensor $\mathbf{F_{G}^j}$ for $\mathbf{M_{G_j}}$. 

\subsection{Shape correlation tensor}
\begin{figure}[h]
\centering
\includegraphics[width=7cm,height=4cm]{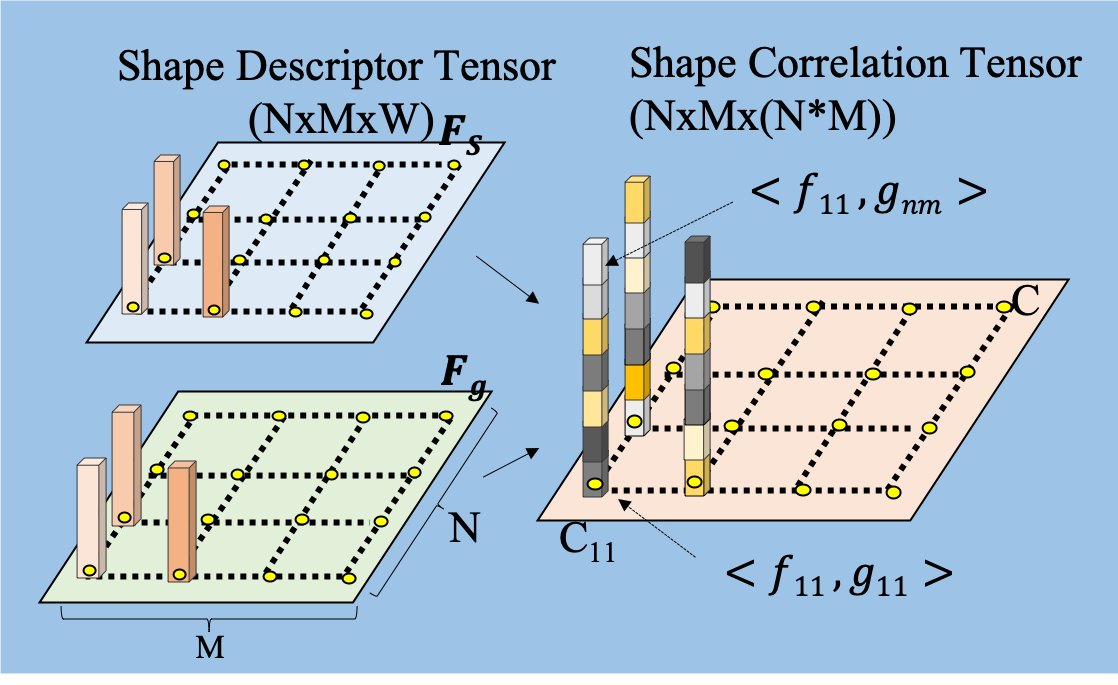}
\caption{The schema of formulating correlation tensor process.}
\label{fig1235}
\end{figure}

As shown in Figure \ref{fig1235}, for the two source and target grid points $\mathbf{M_{S_i}}$ and $\mathbf{M_{G_j}}$ with shape descriptor tensors $\mathbf{F^i_{S}}=[\mathbf{f_{ij}}=\mathcal{U}(\mathbf{x_{ij}},\mathbf{S_i})]$ and $\mathbf{F^i_{G}}=[ \mathbf{g_{ij}}=\mathcal{U}(\mathbf{x_{ij}},\mathbf{G_i})]$, our next step is to define the shape correlation tensor between the input and target shape descriptor tensors. We define the shape correlation tensor in the following step. Let $\mathcal{M}$ be a similarity metric, such that $\mathcal{M}: \mathbb{R}^d \times \mathbb{R}^d \to \mathbb{R}$. In this paper, we simply let $\mathcal{M}$ as inner product. $\forall \mathbf{f_{ij}} \in \mathbf{F^i_{S}}$, we sort the its point-wise correlation with elements in $\mathbf{F^j_{G}}$ as $\mathbf{C_{ij}} \in \mathbb{R}^{t}$ and $t=nm$, where\\

\begin{equation}
\mathbf{C_{ij}}=[\mathcal{M}(\mathbf{f_{ij}},\mathbf{g_{11}}),\mathcal{M}(\mathbf{f_{ij}},\mathbf{g_{12}}),...,\mathcal{M}(\mathbf{f_{ij}},\mathbf{g_{md}})]
\end{equation}\\

We define $\mathbf{C}=[\mathbf{C_{ij}}] \in \mathbb{R}^{n\times m \times t}$ as the shape correlation tensor. It has t-dimensional channel to save the correlation information between each the point in $\mathbf{M_{S_i}}$ with all the points in $\mathbf{M_{G_j}}$. We normalize each channel of element $\mathbf{C_{ij}}$ in the shape correlation tensor. 

\subsection{Shape transformation prediction}
Before we discuss shape transformation prediction, we firstly review two classical parametric functions for rigid and non-rigid transformations. For affine transformation including translation, scaling, rotation and shear. Let $\theta_{rigid} = \{ \alpha, r_1,r_2,r_3,r_4, s_1,s_2\}$ and we have\\ 

$T_{\theta}$$=\begin{bmatrix}
 r_1 \cos \alpha& r_2\sin \alpha & s_1 \\
 r_4 \sin \alpha& r_5 \cos \alpha &s_2 \\
 0 &0 &1
\end{bmatrix}$$\\ \\
$
\noindent Even though we do not discuss the rigid case in this paper, our model can be easily adjusted for rigid registration. 

For non-rigid transformation, let $\mathbf{\theta_{nonrigid}}$ be the controlling points in Thin Plate Spine. In this paper, we choose 9/27 controlling points distributed as a $3\times3$/$3\times3\times3$ grid for 2D/3D data. For a pair of 2D source and target point sets, our target $\mathbf{\theta_{nonrigid}}=\{ (\theta_1,\theta_2),...,(\theta_{17},\theta_{18}) \}$, are a set of coordinates of nine controlling points in TPS \cite{bookstein1989principal}. Let the original controlling points in TPS be $\theta_0$ and $\theta_0=[(0,0),(-1,0),...,(1,-1)]$. For a pair of 3D source and target point sets, our target $\mathbf{\theta_{nonrigid}}=\{ (\theta_1,\theta_2,\theta_3),...,(\theta_{79},\theta_{80}, \theta_{81})\}$, are a set of coordinates of nine controlling points in TPS \cite{bookstein1989principal}. Let the original controlling points in TPS be $\theta_0$ and $\theta_0^{2D}=[(0,0),(-1,0),...,(1,-1)]$ and $\theta_0^{3D}=[(0,0,0),(-1,0,0),(1,0,0),  ... ,  (1,1,1)].$ After achieving new positions of controlling points $\mathbf{\theta_{nonrigid}}$, together with $\mathbf{\theta_0}$, we can solve the non-rigid transformation $T_{\theta}$ according to TPS. In this case, we have 18/81 parameters to be optimized for defining the non-rigid transformation to align the 2D/3D source and target point sets. For a given pair of source point set $\mathbf{S_i}$ and target point set $\mathbf{G_j}$ as inputs, based on their shape correlation tensor $\mathbf{C}$ from the previous step, we further use 2D-CNN/3D-CNN with a successive of fully connected layers to predict the desired parameters $\mathbf{\theta}$ in transformation $T_{\theta}$. 

\subsection{From statistical alignment to loss functions}
The last step is to define the loss function between the transformed source point set $T_{\theta}(\mathbf{S_i})$ and the target point set $\mathbf{G_j}$. Due to the disorderliness of point cloud, there is no direct corresponding relationship between these two point sets. Therefore, a distance metric between two point sets instead of point/pixel-wise loss used in image registration should be desired. Besides, the suitable metric should be differentiable and efficient to compute. For 3D point set generation, Fan et al. \cite{fan2017point} first proposed Chamfer Distance loss, which is widely used in practice. Registration problem can be treated as statistical alignment between two distributions of source and target point sets. We treat target point set as centroids of a Gaussian Mixture Model and we fit the transformed source point set as data into this GMM model so that we can maximize the likelihood of the GMM. \\

\noindent \textbf{Chamfer Distance (C.D.).} 
Chamfer loss is a simple and effective metric to be defined on two non-corresponding point sets. It dose not require the same number of points and in many tasks and it provides high quality results in practice. We define the Chamfer loss on our transformed source point set $T_{\theta}(\mathbf{S})$ and target points set $\mathbf{G} $ as:\\

\begin{equation} 
\begin{split}L_{\text{Chamfer}}(T_{\theta}(\mathbf{S}),\mathbf{G}|\gamma)
 &= \sum_{x\in T_{\theta}(\mathbf{S})}\min_{y \in \mathbf{G}}||x-y||^2_2\\
 &+ \sum_{y\in \mathbf{G}}\min_{x \in T_{\theta}(\mathbf{S})}||x-y||^2_2
\end{split}
\end{equation}\\

where $\gamma$ represents all the parameters in MLP layers and 2D-CNN layers from section 2 B, 2 C and 2 D. In this paper, we use Chamfer Distance (C.D.) as evaluation metric.\\

\noindent \textbf{Gaussian Mixture Model (GMM) loss.} Let our source point set $\mathbf{S}=(\mathbf{x_1},\mathbf{x_2},..., \mathbf{x_N})$ and transformed target point set $\mathbf{T_{\theta}(\mathbf{S})}=(T_{\theta}(\mathbf{x_1}),T_{\theta}(\mathbf{x_2}),..., T_{\theta}(\mathbf{x_N}))$. The target point set is $\mathbf{G}=(\mathbf{y_1},\mathbf{y_2},..., \mathbf{y_M})$ where $\mathbf{x_i} \text{ and } \mathbf{y_i} \in \mathbb{R}^2/\mathbb{R}^3$ in our paper. We consider Gaussian-mixture model $p(T_{\theta}(\mathbf{x_i})) = \sum_{m=1} ^M \frac{1}{M}p(T_{\theta}(\mathbf{x_i})|m)$ with $\mathbf{x}|m\sim N(\mathbf{y_m, \sigma^2 \mathbf{I}_2})$, where our target point set acts as the 2/3-dimensional centroids of equally-weighted Gaussian mixture model. In general we want our predicted point set to maximally satisfy the Gaussian Mixture model. Therefore, we define the loss function (GMM loss) as :\\

\begin{equation}
L_{\text{GMM}}(T_{\theta}(\mathbf{S}),\mathbf{G}|\gamma)= - \sum_{\mathbf{x} \in \mathbf{S}} \log \sum_{\mathbf{y} \in \mathbf{G}} e^ {- \frac{1}{2} \left\|\frac{T_{\theta}(\mathbf{x})-\mathbf{y}}{\sigma}\right\|^2}
\end{equation}\\

,where $\gamma$ represents all the parameters in MLP layers and 2D-CNN layers from section 2 B, 2 C, and 2 D. $\sigma$ is the standard deviation in GMM. We set $\sigma$ to be identical for each Gaussian distribution in GMM. $\sigma$ is a hyper-parameter to choose in practice. Even though it is a constant for each input, we have more sophisticated strategy for choosing it in practice as discussed in section 2 F. We use GMM loss as our loss function in this paper. 

\subsection{Model settings}
We train our network using batch data form training data set $\{(\mathbf{S^i},\mathbf{G^i}) | (\mathbf{S^i}, \mathbf{G^i}) \in \mathbf{D} \}_{i=1,2,...,b}$. b is the batch size and is set to 16. For learning the shape descriptor tensor in 2 B, the input is $N\times4/N\times6$ matrix and we use 4 MLP layers with dimensions (16,32,64,128) and a Maxpool layer to convert it to a 128-dimensional descriptor for each point in $11 \times 11$ reference grid. For the shape correlation tensor $\mathbf{C}$ discussed in 2 C and 2 D, we use three 2D-CNN/3D-CNN layers with kernel size (3,3),(4,4),(5,5) and dimension (128,256,512) with two successive fully connected layers with dimensions (64, 18)/(512,81). Learning rate is set as 0.0001 with 0.995 exponential decay with Adam optimizer. We use leaky-ReLU \cite{xu2015empirical} activation function and implement batch normalization \cite{ioffe2015batch} for every layer except the output layer. We use deterministic annealing for the standard deviation $\sigma$ which is initially set to 1, and for each step $n$ we reduce it to $\sqrt{1/n}$ until a margin value of 0.1. Gradual reducing $\sigma$ leads to a coarse-to-fine match. For outlier and missing points case, we slightly increase the margin value to 0.12.
\begin{table*}
\begin{center}
\begin{tabular}{ccc}
\hline
Methods&CD&Time 
\\
\hline
CPD (Train) \cite{myronenko2007non} &$0.0038\pm0.0031$&$\sim$ 12 $\text{ hours}$\\
PR-Net (Train) &0.0037$\pm$0.0014&$\sim$ 13 $\text{ minutes}$\\
CPD (test) \cite{myronenko2007non} &0.0038$\pm$0.0032& $\sim$ 12 $\text{ hours}$\\
PR-Net (Test) &0.0044$\pm$0.0016&$\sim$ 8 $\text{ seconds}$\\
\hline
\end{tabular}
\end{center}
\caption{Performance comparison with CPD for registering $10k$ pairs of point sets at deformation level $0.5$.}
\label{t1}
\end{table*}

\begin{table*}
\begin{center}
\begin{tabular}{cc}
\hline
Deform. Level&Chamfer Distance
\\
 \hline
 0.2&0.0013$\pm$0.0005\\
0.3&0.0019$\pm$0.0008\\
0.8&0.0161$\pm$0.0057\\
1.0&0.0153$\pm$0.0052\\
1.5&0.1267$\pm$0.0872\\
\hline
\end{tabular}
\end{center}
\caption{Quantitative testing performance for 2D fish shape point set registration at different deformation level (Deform. Level)}
\label{tt2}
\end{table*}

\section{Experiments}
In this section, we implement a set of experiments to validate the performance of our proposed PR-Net for non-rigid point set registration from different aspects (i.e. accuracy and time). In section 3 A, we discuss how we prepare the experimental dataset. In section 3 B, we compare PR-Net with non-learning based non-rigid point set registration method. In section 3 C, we validate the robustness of PR-Net against the different level of geometric deformation. In section 3 D, we validate the robustness of PR-Net against the different types of noise. In section 3 E, we further verify that PR-Net can handle registration tasks for various types of dataset.  

\subsection{Dataset preparation}
The point cloud data is often featured with geometric structural variations with presence of a variety of noise (e.g. outliers, missing points), which poses challenges for point set registration. An effective registration solution should be robust to the presence of those noise to provide the desired geometric transformation. Therefore, in order to assess PR-Net's performance, we simulate the commonly recognized noise to the raw point sets to prepare the experimental data. To prepare the geometric structural variation, we randomly choose a certain number of samples from the point set and use them as the controlling points of a thin plate spline (TPS) transformation. A zero-mean Gaussian is superposed to each controlling point to simulate a random drift from their original positions. The TPS is then applied to synthesize the deformed point set with different level of structural variation. The $1/2$ of standard deviation of the above mentioned Gaussian is used to measure the deformation level. To prepare the position drift (P.D.) noise, we applied a zero-mean Gaussian to each sample from the point set. The level of P.D. noise is defined as the standard deviation of Gaussian. To prepare the data incompleteness (D.I.) noise, we randomly remove a certain amount of points from the entire point set. The level of D.I. noise is defined as ratio of the eliminated points and the entire set. To prepare the data outlier (D.O.) noise, we randomly add a certain amount of points generated by a zero-mean Gaussian to the point set. The level of D.O. noise is defined as the ratio of the added points to the entire point set. For all tests, we use the Chamfer Distance (C.D.) between a pair of point sets to provide a quantitative score to evaluate the registration performance.
\begin{figure*}
\centering
\includegraphics[width=17cm,height=14.5cm]{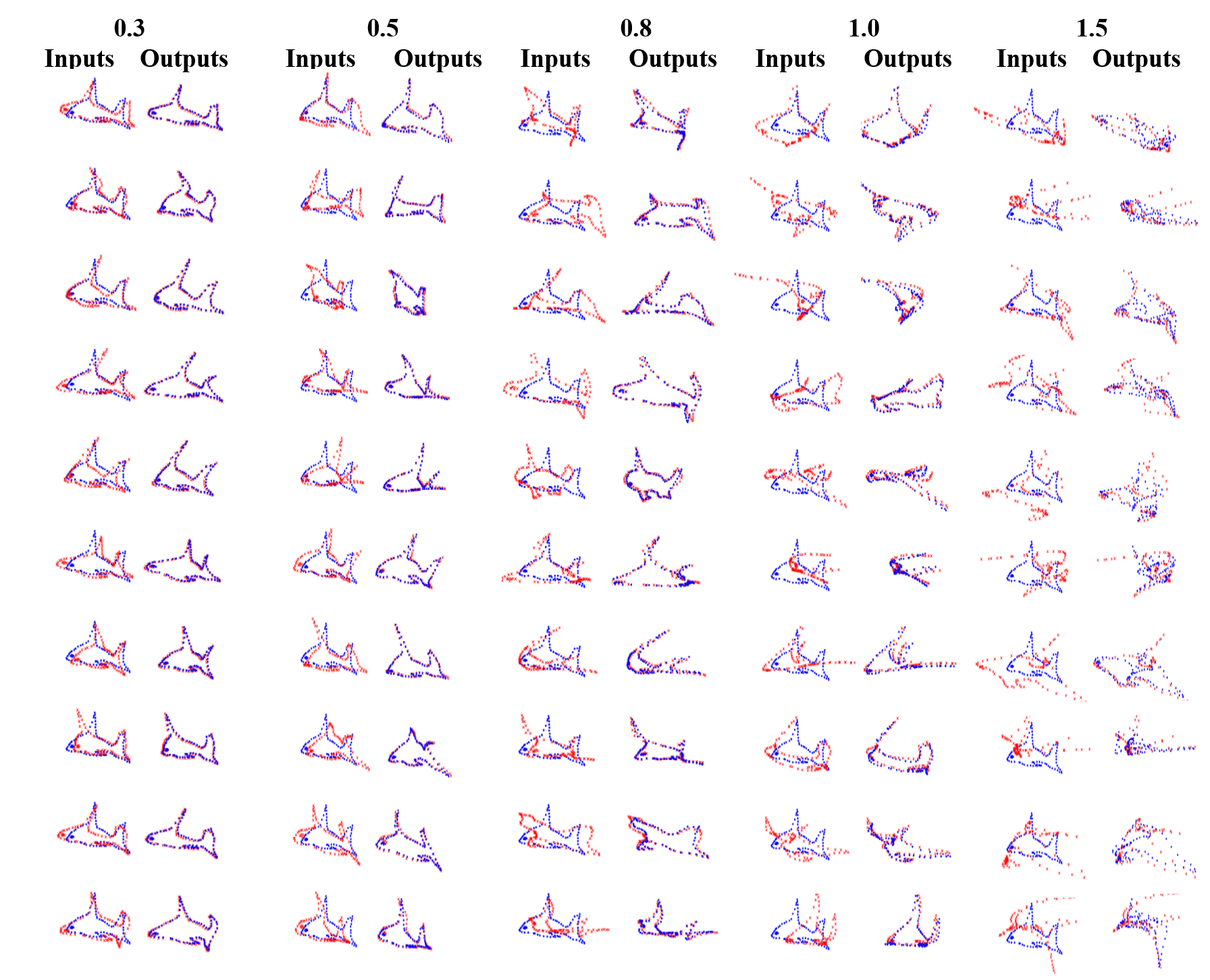}
\caption{Testing results for 2D fish shape point set registration at different deformation levels. The deformation level increases from 0.3 to 1.5 from left to right. The presented shapes are randomly selected from same testing batch. The blue shapes are source point sets and the red shapes are target point sets. Please zoom-in for better visualization.}
\label{fnn1}
\end{figure*}

\subsection{Comparison to Non-learning based Approach}
Different from previous efforts, the proposed PR-Net
is a learning-based non-rigid point set registration method, which can learn the registration pattern to directly predict the non-parametric geometric transformation for the point sets alignment. As a learning-based approach to predict the non-rigid registration, it is not applicable to have a direct
comparison between PR-Net and other existing non-rigid iterative registration methods. To compare our method to non-learning based iterative method (i.e. Coherent Point Drift (CPD) \cite{myronenko2007non}), we design the experiment as follows to assess both time and accuracy performance. \\

\noindent \textbf{Experimental Setup:} We conduct tests to compare PR-Net with the non-learning based approach. Coherent Point Drifts (CPD)\cite{myronenko2007non} is a highly recognized non-rigid point set registration method. In this test, we synthesize 2D deformed fish data with deformation level of $0.5$ to prepare $10k$ training dataset and $10k$ testing dataset. Our PR-Net is firstly trained before applied to the $10k$ testing dataset. The CPD is directly applied to the $10k$ testing dataset.\\

\noindent \textbf{Result:} We list the experimental result in the table \ref{t1}. The second column shows the mean and standard deviation of all $10k$ C.D. after registration. The third column shows the time used for registering the $10k$ pairs of point sets. As we expect, after training PR-Net can perform the real-time non-rigid point set registration. The time used to register $10k$ pairs of point sets is around $8$ seconds, which is order of magnitude less than the time ($12$ hours) consumed by CPD for point set registration of the entire $10k$ dataset. This is because of the fact that CPD needs to repeatedly start over a new iterative process for a new pair of point sets. PR-Net clearly gains advantage over the non-learning based method by providing a faster solution to non-rigid point registration. We also want to note that it takes around $13$ minutes to train our PR-Net on the $10k$ dataset with a comparative performance, which is also significantly less than $22$ hours used by CPD.

In addition to the efficiency (registration speed), we are also interested in the effectiveness that indicates how well PR-Net can generalize from training data to directly predict the desired geometric transformation for non-rigid point set registration. The comparative training and testing C.D. results are listed on the second column. The small difference between training and testing C.D. indicates a comparative small performance degradation from training to testing. Furthermore, we notice that C.D. of PR-Net has a smaller standard deviation than that of CPD, which suggests that PR-Net can provide a more stable registration as it obtains generalization ability to adapt properly to previously unseen data. In contrast, the CPD treats every new pair of point sets independently and has to repeatedly register them from the start.

\subsection{Robust to Geometric Deformation}
In this experiment, we take a detailed investigation on how well the PR-Net performs point set registration for 2D shapes at different deformation levels. This experiment shows a basic assessment of our model's performance and capacity for registering unseen highly deformed testing shapes. \\

\noindent \textbf{Experimental Setup:} We conduct tests to verify how well PR-Net performs on the data with different levels of geometric deformation. In this test, we synthesize 2D deformed fish data with deformation levels from $0.3$ to $1.5$ to cover a good range of shape structural variation. The deformed 2D fish shapes are shown in Figure \ref{fnn1}. For each level of deformation, we simulate $20k$ point sets as target point sets for training and simulate additional $10k$ point sets for testing. The quantitative result is shown in Table \ref{tt2}. \\

\noindent \textbf{Result:} After training, the PR-Net is applied to register testing datasets with different deformation levels. The quantitative experimental results are listed in Table \ref{tt2}. The second column lists the C.D. scores for a registered pair of source and target point sets with different deformation levels. As we can see from the evaluation, PR-Net can achieve impressive performance on non-rigid point set registration when the deformation level is less than $1.0$ and the Chamfer Distance remains as low as $0.0153$ as shown in Table \ref{tt2}. However when the deformation level reaches $1.5$, there is a huge jump of C.D. from $0.0153$ to $0.1267$. This indicates that our model's registration capacity dose have a clear upper bond. Once the deformation level reaches or higher than this upper bond, the performance of PR-Net can be dramatically reduced. We further check the qualitative results for better understanding PR-Net's performance.
\begin{table*}
\begin{center}
\begin{tabular}{||c|c||c|c||c|c||c|c||}
\hline
P.D. Level&C.D.&D.O. Level&C.D.&D.I. Level&C.D.
\\
 \hline
 0.05&0.0052$\pm$0.0009& 0.05&0.003$\pm$0.001&0.05&0.0134$\pm$0.0038 \\
0.08&0.0074$\pm$0.001&0.15&0.0033$\pm$0.001&0.2&0.0147$\pm$0.0053\\
0.1&0.0093$\pm$0.0012&0.25&0.0088$\pm$0.0029&0.3&0.0154$\pm$0.0053\\
0.15&0.0145$\pm$0.002&0.3&0.0103$\pm$0.003&0.45&0.0178$\pm$0.0053\\
0.2&0.0204$\pm$0.0029&0.5&0.0195$\pm$0.0061&0.6&0.021$\pm$0.0067\\
\hline
\end{tabular}
\end{center}
\caption{Quantitative testing performance for 2D fish shape point set registration at different deformation level $0.5$ in presence of various noise such as Point Drift (P.D) noise, Data Outlier (D.O.) noise, and Data Incompleteness (D.I.) noise. }
\label{t2}
\end{table*}

\begin{figure*}
\centering
\includegraphics[width=11cm,height=10cm]{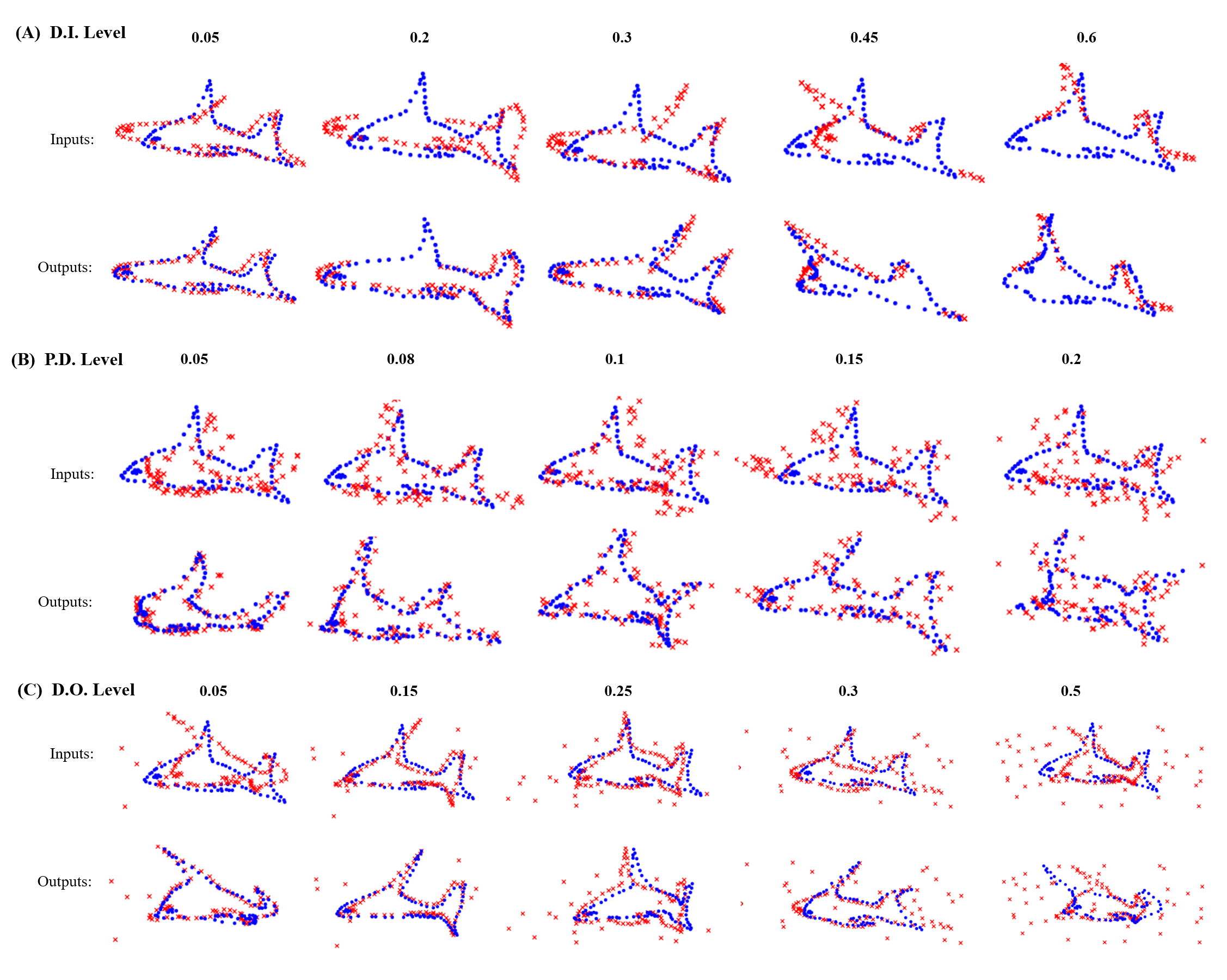}
\caption{Testing results for 2D fish shape point set registration at deformation level $0.5$ in presence of various noise. (A) Performance in presence of Data Incompleteness (D.I.) noise. (B) Performance in presence of Point Drift (P.D.) noise. (C) Performance in presence of Data Outlier (D.O.) noise. Blue shapes are source point sets and red ones are target point sets. Please zoom-in for better visualization.}
\label{fn}
\end{figure*}
The corresponding qualitative results are demonstrated in Figure \ref{fnn1}, which illustrates the pairs of point sets before and after registration. From the Figure \ref{fnn1}, we can clearly see that the transformed source point set (in blue color) structurally aligns well with the target point set (in red color), which verifies PR-Net's registration capacity. Especially when deformation level is equal or less then $1.0$, as shown in \ref{fnn1}, PR-Net almost perfectly aligns the source and target point sets. As we mentioned before, when the deformation level reaches $1.5$, the quantitative result experiences a dramatic drop. As displayed in Figure \ref{fnn1}, for this deformation level $1.5$, the geometric structure of 2D fish is significantly deteriorated, which poses much more challenges in determining the desired geometric transformation. Even for human beings, it is hard to tell the geometric meaning of the target point sets (Red shapes in Figure \ref{fnn1}). But this also indicates that TPS, as a parametric geometric transformation model, might be limited in modeling the large structural variation in our test. We further investigate more complex geometric transformation model or model-free geometric transformation in our separate research reports.
\begin{table*}
\begin{center}
\begin{tabular}{lcccc}
\hline
Deform. Level&0.3&0.5&0.8&1.0 \\
 \hline
Hand &0.0013$\pm$0.0006&0.0025$\pm$0.0013&0.0056$\pm$0.0025 &0.0105$\pm$0.0047  \\
Skeleton &0.0012$\pm$0.0005&0.0022$\pm$0.0010&0.0081$\pm$0.0049 &0.0087$\pm$0.0047  \\
Skull &0.0017$\pm$0.0008&0.0029$\pm$0.0011&0.0052$\pm$0.0022 & 0.01$\pm$0.0036\\
\hline
\end{tabular}
\end{center}
\caption{Quantitative testing performance for skull, hand, and skeleton 2D shapes at different deformation level from 0.3 to 1.0.}
\label{t3}
\end{table*}
\begin{figure*}
\centering
\includegraphics[width=7.5cm,height=9cm]{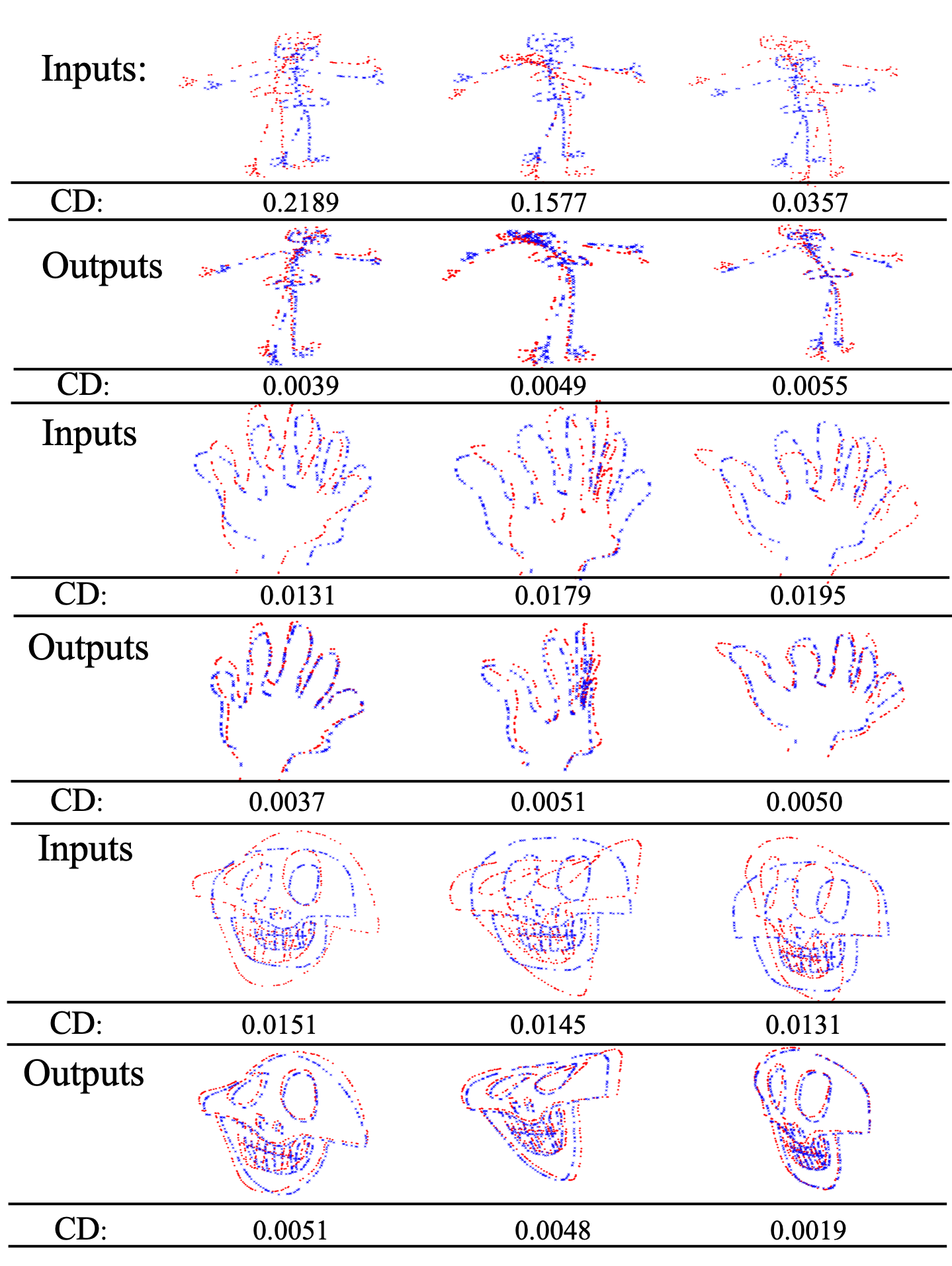}
\caption{Testing performance for skull, hand and human skeleton shapes.Blue shapes are source point sets and red ones are target point sets. Please zoom-in for better visualization. The corresponding C.D. for each input and output pair is presented below it.}
\label{other}
\end{figure*}
\subsection{Robust to Data Noise}
While using the sensors such as LIDAR sensor and laser scanner, it is unavoidable that the data might be acquired with a variety types of noises. An effective non-rigid registration method should be robust to those noise in addition to the structural variations as discussed in previous section. Therefore, in this section, we focus on testing how well PR-Net can predict the non-rigid registration from the noisy dataset. \\

\noindent \textbf{Experimental Setup:} In this experiment, we carry out a set of tests to validate PR-Net's performance against different types of data noise including P.D. noise, D.I. noise, and D.O. noise. We simulate the noisy data through introducing three types of noise with five different levels to the target point set at deformation level of $0.5$. The level of noise is defined in the section of data preparation. The Figure \ref{fn} illustrates the noisy target point set (in red color) in contrast to the source point set (in blue color). The quantitative result is demonstrated in Table \ref{t2}.\\

\noindent \textbf{Result:} Figure \ref{fnn1} demonstrates the PR-Net's performance with clean data for comparison. Given the source point set (in red color) and target point set (in blue color), PR-Net succeeds in transforming source point set to align with the target one for the clean data.

For investigating PR-Net's performance on noise data, in Figure \ref{fn} (A), we apply D.I. noise to target point set by increasingly removing point samples as shown from left to right in a row. The registration results show that our PR-Net is capable of robustly aligning the source point set (red) with target (blue) in this condition. Even for the situation when D.I. noise level is $0.6$ and the majority of target shape is missing, PR-Net can still align the remaining parts such as the top and tail of the target fish. An interesting observation is that for the missing parts, even without any target information, the transformed source point sets seem to be natural and preserve the original geometric meaning. For example when the D.I. level reaches $0.6$, the transformed source point sets not only match the targets, but the shape in general still has the geometric meaning for the missing parts and it can be easily recognized as a ``fish" shape. As shown in Table \ref{t2}, the quantitative result shows that C.D. linearly increases when D.I. Level increases from $0.05$ to $0.6$, which indicates PR-Net's high resistance to D.I. Noise.

In Figure \ref{fn} (B), we apply P.D. noise to target point set by increasingly adding Gaussian noise as shown from left to right in a row. As shown in Figure \ref{fn}, though the positions of target point sets are dramatically drifted by Gaussian noise, our PR-Net still effectively predicts the desired geometric transformation. Especially when the P.D. noise level is higher than $0.15$, even though the boundary of the fish shape is dramatically drifted, the transformed source shapes have smooth boundary and acceptable alignment with the target ones. From the quantitative results, as shown in Table \ref{t2}, the C.D. of registered pairs is less than $0.01$ when the P.D. noise level is under $0.15$, which indicates almost perfect alignment. 
\begin{figure*}
\centering
\includegraphics[width=12cm,height=4cm]{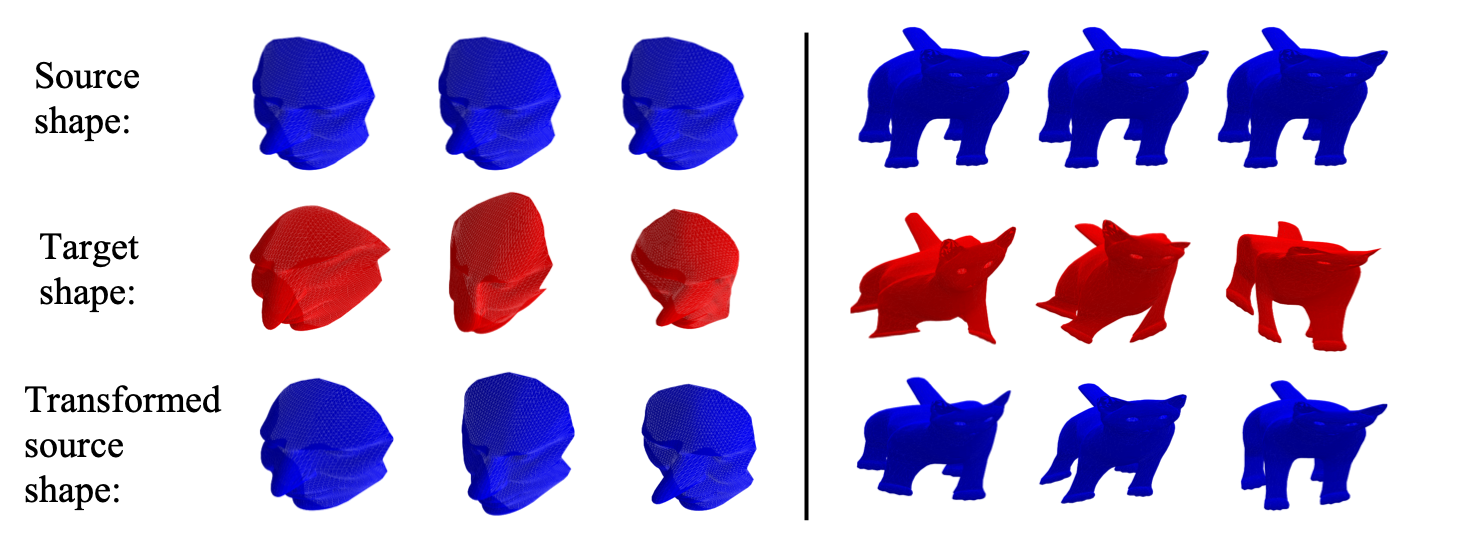}
\caption{Testing registration performance for 3D face and cat point sets. The blue shapes are source shapes and the red shapes are target ones. We plot the mesh of shapes for better visualization.}
\label{3d}
\end{figure*}
The D.O. noises is added to target point set in Figure \ref{fn} (C) as shown from left to right in a row. The registration result demonstrates that the alignment between the source and target shapes is not significantly affected by outlier points in target set when the D.O. noise level is less than $0.3$. The quantitative results show that the C.D. of registered pairs remain as low as $0.0103$ when the D.O noise level is $0.3$. However, when the D.O. noise level reaches as high as $0.5$ the C.D. of registered pairs jumps from $0.0103$ to $0.0195$, which indicates that PR-Net starts suffering dramatic performance degradation affected by the large amount of added outliers. 

\subsection{Results on Data Variety}
In this experiment, we take a further step to investigate how well the PR-Net performs point set registration for other 2D/3D shapes at different deformation levels. We are especially interested in point set registration of non-contour based 2D shapes, as well as 3D shapes since  the  3D  data  have been  gaining  great  attention  in  community  with  recent advancements in 3D acquisition and computation resources.\\

\noindent \textbf{Experimental Setup:} We further conduct tests to verify how well PR-Net performs on the dataset of various shapes and patterns, such as skeleton, skull, hand, face (3D shape) and cat (3D shape). For each type of dataset, with different levels of geometric deformation, we simulate $20k$ point sets as target point sets for training and simulate additional $10k$ point sets for testing. For 3D shapes, we randomly sample points from the mesh data set. While we training PR-Net on 3D shapes, we only sample $512$ out of $20K$ points from an input 3D model and $125(5\times5\times5)$ controlling points for learning descriptor tensor and correlation tensor, which already provided reasonable registration. Due to the computation complexity, there is a clear trade-off between performance with computation efficiency. We randomly select a few samples at deformation level of $0.5$ to visualize the alignment result in Figure \ref{other} and Figure \ref{3d}. We present the quantitative evaluation of registration in Table \ref{t3}, which presents the mean and standard deviation of C.D. between registered pairs.\\

\noindent \textbf{Result:} PR-Net demonstrates robust registration performance for various categories of 2D shapes (e.g. skull, skeleton and hand), based on the selected examples from the testing dataset are demonstrated in Figure \ref{other} and the corresponding quantitative testing results for comparison of these three different shapes are shown in Table \ref{t3}. The decreasement in C.D. values from pre to post registration suggests that PR-Net can successfully align deformable pairs of various shapes. As shown in Figure \ref{other}, for the current deformation level $0.5$, PR-Net shows robust performance regarding to different shapes. There is no obvious difference among the registration results of them. When zooming in for a more detailed observation, the missing registration part can still be noticed such as the upper line of the skull in row 2. As shown in Table \ref{t3}, for comparing the quantitative results on these three shapes, the result on Skull Shape is slightly worse than other two shapes when deformation level is low. But for higher deformation level, the performance on Skull shape becomes comparative to other two shapes. This validates the robust performance of PR-Net towards non-rigid point set registration over a variety of shapes in presence of different geometric deformation level. In Figure \ref{3d}, we demonstrate that PR-Net is applicable for 3D point set registration. As shown in Figure \ref{3d}, for the general part of the target shape, our model can correctly predict the registration transformation to align them. As to aligning the more subtle part of source and target point sets, there is still space to improve PR-Net's performance. The straightforward method to improve the performance is to increase the number of sampling points from surface and as well as the controlling points for learning the shape descriptor tensor with acceptable computation cost. The comparison result across different categories of shapes indicates the consistent performance of PR-Net.  

\section{Conclusion \& Discussion}
This paper introduces a novel learning-based approach to our research community for non-rigid point set registration. In contrast to non-learning based methods (e.g. Coherent Point Drift), the learning based approaches for point set registration are rarely studied (to the best of our knowledge, PR-Net might be the first work that can actually generalize from training to predict geometric transformation for non-rigid point set registration). Possible reasons behind are 1) the irregular format of the point cloud data poses a challenge for standard learnable operator (e.g. discrete convolutions) to operates over non-grid structured data for point feature learning and 2) it is not obvious to define an appropriate geometric transformation to transform source point set to the target one. The PR-Net provides the shape descriptor tensor and correlation tensor for the solution of feature learning, and uses the thin plate spline to model the geometric transformation. Though PR-Net is capable of learning the point registration, there are still some challenges that are left to be addressed. Firstly, our current PR-Net indirectly uses the regular grids to assist with the shape feature learning. A continuous operator, which can directly be applied on point for feature learning, would be more applicable for point registration. Secondly, PR-Net uses the TPS to model the geometric transformation. Though it predicts impressive registration performance for shapes with moderate deformation, the unsatisfactory performance for shapes with large deformation motivates us to study a model-free geometric transformation (e.g. the displacement field). It would also be of great interest to extend PR-Net to other data modality such as 2D Image and 3D volumetric data. We will report those research outcomes in separate papers. 

\newpage

\begin{rezabib}
@article{ma2016non,
  title={Non-rigid point set registration by preserving global and local structures},
  author={Ma, Jiayi and Zhao, Ji and Yuille, Alan L},
  journal={IEEE Transactions on image Processing},
  volume={25},
  number={1},
  pages={53--64},
  year={2016},
  publisher={IEEE}
}

@article{jian2011robust,
  title={Robust point set registration using gaussian mixture models},
  author={Jian, Bing and Vemuri, Baba C},
  journal={IEEE transactions on pattern analysis and machine intelligence},
  volume={33},
  number={8},
  pages={1633--1645},
  year={2011},
  publisher={IEEE}
}

@article{bai2007skeleton,
  title={Skeleton pruning by contour partitioning with discrete curve evolution},
  author={Bai, Xiang and Latecki, Longin Jan and Liu, Wen-Yu},
  journal={IEEE transactions on pattern analysis and machine intelligence},
  volume={29},
  number={3},
  year={2007},
  publisher={IEEE}
}

@article{bai2008path,
  title={Path similarity skeleton graph matching},
  author={Bai, Xiang and Latecki, Longin Jan},
  journal={IEEE transactions on pattern analysis and machine intelligence},
  volume={30},
  number={7},
  pages={1282--1292},
  year={2008},
  publisher={IEEE}
}

@article{myronenko2009image,
  title={Image registration by minimization of residual complexity},
  author={Myronenko, Andriy and Song, Xubo},
  year={2009},
  publisher={IEEE}
}

@inproceedings{wu2012online,
  title={Online robust image alignment via iterative convex optimization},
  author={Wu, Yi and Shen, Bin and Ling, Haibin},
  booktitle={2012 IEEE Conference on Computer Vision and Pattern Recognition},
  pages={1808--1814},
  year={2012},
  organization={IEEE}
}

@article{ma2014robust,
  title={Robust point matching via vector field consensus.},
  author={Ma, Jiayi and Zhao, Ji and Tian, Jinwen and Yuille, Alan L and Tu, Zhuowen},
  journal={IEEE Trans. image processing},
  volume={23},
  number={4},
  pages={1706--1721},
  year={2014}
}
@inproceedings{ling2005deformation,
  title={Deformation invariant image matching},
  author={Ling, Haibin and Jacobs, David W},
  booktitle={Computer Vision, 2005. ICCV 2005. Tenth IEEE International Conference on},
  volume={2},
  pages={1466--1473},
  year={2005},
  organization={IEEE}
}
@inproceedings{klaus2006segment,
  title={Segment-based stereo matching using belief propagation and a self-adapting dissimilarity measure},
  author={Klaus, Andreas and Sormann, Mario and Karner, Konrad},
  booktitle={Pattern Recognition, 2006. ICPR 2006. 18th International Conference on},
  volume={3},
  pages={15--18},
  year={2006},
  organization={IEEE}
}

@article{maintz1998survey,
  title={A survey of medical image registration},
  author={Maintz, JB Antoine and Viergever, Max A},
  journal={Medical image analysis},
  volume={2},
  number={1},
  pages={1--36},
  year={1998},
  publisher={Elsevier}
}

@inproceedings{besl1992method,
  title={Method for registration of 3-D shapes},
  author={Besl, Paul J and McKay, Neil D},
  booktitle={Sensor Fusion IV: Control Paradigms and Data Structures},
  volume={1611},
  pages={586--607},
  year={1992},
  organization={International Society for Optics and Photonics}
}
@inproceedings{raguram2008comparative,
  title={A comparative analysis of RANSAC techniques leading to adaptive real-time random sample consensus},
  author={Raguram, Rahul and Frahm, Jan-Michael and Pollefeys, Marc},
  booktitle={European Conference on Computer Vision},
  pages={500--513},
  year={2008},
  organization={Springer}
}
@article{yuille1988computational,
  title={A computational theory for the perception of coherent visual motion},
  author={Yuille, Alan L and Grzywacz, Norberto M},
  journal={Nature},
  volume={333},
  number={6168},
  pages={71},
  year={1988},
  publisher={Nature Publishing Group}
}
@book{sonka2014image,
  title={Image processing, analysis, and machine vision},
  author={Sonka, Milan and Hlavac, Vaclav and Boyle, Roger},
  year={2014},
  publisher={Cengage Learning}
}

@article{tam2013registration,
  title={Registration of 3D point clouds and meshes: a survey from rigid to nonrigid.},
  author={Tam, Gary KL and Cheng, Zhi-Quan and Lai, Yu-Kun and Langbein, Frank C and Liu, Yonghuai and Marshall, David and Martin, Ralph R and Sun, Xian-Fang and Rosin, Paul L},
  journal={IEEE transactions on visualization and computer graphics},
  volume={19},
  number={7},
  pages={1199--1217},
  year={2013},
  publisher={Institute of Electrical and Electronics Engineers, Inc., 3 Park Avenue, 17 th Fl New York NY 10016-5997 United States}
}

@inproceedings{chui2000new,
  title={A new algorithm for non-rigid point matching},
  author={Chui, Haili and Rangarajan, Anand},
  booktitle={Computer Vision and Pattern Recognition, 2000. Proceedings. IEEE Conference on},
  volume={2},
  pages={44--51},
  year={2000},
  organization={IEEE}
}

@article{bookstein1989principal,
  title={Principal warps: Thin-plate splines and the decomposition of deformations},
  author={Bookstein, Fred L.},
  journal={IEEE Transactions on pattern analysis and machine intelligence},
  volume={11},
  number={6},
  pages={567--585},
  year={1989},
  publisher={IEEE}
}

@article{ma2015robust,
  title={Robust L2E Estimation of Transformation for Non-Rigid Registration.},
  author={Ma, Jiayi and Qiu, Weichao and Zhao, Ji and Ma, Yong and Yuille, Alan L and Tu, Zhuowen},
  journal={IEEE Trans. Signal Processing},
  volume={63},
  number={5},
  pages={1115--1129},
  year={2015}
}

@inproceedings{wang2016path,
  title={Path following with adaptive path estimation for graph matching},
  author={Wang, Tao and Ling, Haibin},
  booktitle={Thirtieth AAAI Conference on Artificial Intelligence},
  year={2016}
}

@inproceedings{zhou2015multi,
  title={Multi-image matching via fast alternating minimization},
  author={Zhou, Xiaowei and Zhu, Menglong and Daniilidis, Kostas},
  booktitle={Proceedings of the IEEE International Conference on Computer Vision},
  pages={4032--4040},
  year={2015}
}

@inproceedings{su2015multi,
  title={Multi-view convolutional neural networks for 3d shape recognition},
  author={Su, Hang and Maji, Subhransu and Kalogerakis, Evangelos and Learned-Miller, Erik},
  booktitle={Proceedings of the IEEE international conference on computer vision},
  pages={945--953},
  year={2015}
}

@inproceedings{sharma2016vconv,
  title={Vconv-dae: Deep volumetric shape learning without object labels},
  author={Sharma, Abhishek and Grau, Oliver and Fritz, Mario},
  booktitle={Computer Vision--ECCV 2016 Workshops},
  pages={236--250},
  year={2016},
  organization={Springer}
}
@inproceedings{maturana2015voxnet,
  title={Voxnet: A 3d convolutional neural network for real-time object recognition},
  author={Maturana, Daniel and Scherer, Sebastian},
  booktitle={Intelligent Robots and Systems (IROS), 2015 IEEE/RSJ International Conference on},
  pages={922--928},
  year={2015},
  organization={IEEE}
}

@inproceedings{verma2018feastnet,
  title={FeaStNet: Feature-Steered Graph Convolutions for 3D Shape Analysis},
  author={Verma, Nitika and Boyer, Edmond and Verbeek, Jakob},
  booktitle={CVPR 2018-IEEE Conference on Computer Vision \& Pattern Recognition},
  year={2018}
}

@inproceedings{masci2015geodesic,
  title={Geodesic convolutional neural networks on riemannian manifolds},
  author={Masci, Jonathan and Boscaini, Davide and Bronstein, Michael and Vandergheynst, Pierre},
  booktitle={Proceedings of the IEEE international conference on computer vision workshops},
  pages={37--45},
  year={2015}
}

@article{qi2017pointnet,
  title={Pointnet: Deep learning on point sets for 3d classification and segmentation},
  author={Qi, Charles R and Su, Hao and Mo, Kaichun and Guibas, Leonidas J},
  journal={Proc. Computer Vision and Pattern Recognition (CVPR), IEEE},
  volume={1},
  number={2},
  pages={4},
  year={2017}}

@inproceedings{zeng20173dmatch,
  title={3dmatch: Learning local geometric descriptors from rgb-d reconstructions},
  author={Zeng, Andy and Song, Shuran and Nie{\ss}ner, Matthias and Fisher, Matthew and Xiao, Jianxiong and Funkhouser, Thomas},
  booktitle={Computer Vision and Pattern Recognition (CVPR), 2017 IEEE Conference on},
  pages={199--208},
  year={2017},
  organization={IEEE}
}

@inproceedings{myronenko2007non,
  title={Non-rigid point set registration: Coherent point drift},
  author={Myronenko, Andriy and Song, Xubo and Carreira-Perpin{\'a}n, Miguel A},
  booktitle={Advances in Neural Information Processing Systems},
  pages={1009--1016},
  year={2007}
}

@inproceedings{rocco2017convolutional,
  title={Convolutional neural network architecture for geometric matching},
  author={Rocco, Ignacio and Arandjelovic, Relja and Sivic, Josef},
  booktitle={Proc. CVPR},
  volume={2},
  year={2017}
}

@inproceedings{balakrishnan2018unsupervised,
  title={An Unsupervised Learning Model for Deformable Medical Image Registration},
  author={Balakrishnan, Guha and Zhao, Amy and Sabuncu, Mert R and Guttag, John and Dalca, Adrian V},
  booktitle={Proceedings of the IEEE Conference on Computer Vision and Pattern Recognition},
  pages={9252--9260},
  year={2018}
}

@article{xu2015empirical,
  title={Empirical evaluation of rectified activations in convolutional network},
  author={Xu, Bing and Wang, Naiyan and Chen, Tianqi and Li, Mu},
  journal={arXiv preprint arXiv:1505.00853},
  year={2015}
}

@article{fan2016point,
  title={A point set generation network for 3d object reconstruction from a single image},
  author={Fan, Haoqiang and Su, Hao and Guibas, Leonidas},
  journal={arXiv preprint arXiv:1612.00603},
  year={2016}
}
@inproceedings{fan2017point,
  title={A Point Set Generation Network for 3D Object Reconstruction from a Single Image.},
  author={Fan, Haoqiang and Su, Hao and Guibas, Leonidas J},
  booktitle={CVPR},
  volume={2},
  number={4},
  pages={6},
  year={2017}
}

@inproceedings{ioffe2015batch,
  title={Batch normalization: Accelerating deep network training by reducing internal covariate shift},
  author={Ioffe, Sergey and Szegedy, Christian},
  booktitle={International Conference on Machine Learning},
  pages={448--456},
  year={2015}
}

\end{rezabib}

\bibliographystyle{IEEEtran}

\end{document}